\documentclass[pra,showpacs,showkeys,amsfonts,amsmath,twocolumn]{revtex4}
\usepackage{bm}
\usepackage{graphicx}
\RequirePackage{mathptm}
\newtheorem{thm}{Theorem}
\newtheorem{cor}{Corollary}

\numberwithin{equation}{section}
\newcommand{\si}[1]{\sigma_{#1}}

\newcommand{\ip}[2]{\langle \,{#1},\,{#2}\,\rangle}

\newcommand{\ro}{\rho}

\newcommand{\La}{\Lambda}

\newcommand{\I}{\mathbb I}
\newcommand{\ket}[1]{|{#1}\rangle}
\newcommand{\bra}[1]{\langle {#1} |}
\newcommand{\cH}{{\mathcal H}}
\newcommand{\C}{\mathbb C}
\newcommand{\R}{\mathbb R}

\newcommand{\fE}{\mathcal E}
\newcommand{\tr}{\mathrm{tr}\,}

\newcommand{\tl}[1]{\boldsymbol #1}
\newcommand{\mr}[1]{\mathrm{#1}}

\newcommand{\DS}{\displaystyle}
\begin{document}
\title{CHSH violation and entropy -- concurrence plane}
\author{{\L}ukasz Derkacz}
\affiliation{Institute of Theoretical Physics\\ University of
Wroc{\l}aw\\
Pl. M. Borna 9, 50-204 Wroc{\l}aw, Poland}
\author{Lech Jak{\'o}bczyk\footnote{
E-mail addres: ljak@ift.uni.wroc.pl}}
\affiliation{Institute of Theoretical Physics\\ University of
Wroc{\l}aw\\
Pl. M. Borna 9, 50-204 Wroc{\l}aw, Poland}
\begin{abstract}
We characterize violation of CHSH inequalities for mixed two-qubit
states by their  mixedness and entanglement. The class of states
that have maximum degree of CHSH violation for a given linear
entropy is also constructed.
\end{abstract}
 \maketitle
\section{Introduction}
Physically allowed degree of entanglement and mixture for two -
qubit mixed states were investigated by Munro \textit{et al.} in
terms of concurrence  $C(\ro)$ and  normalized linear entropy
$S_{\mr{L}}(\ro)$ \cite{MJWK}. These authors characterize the subset
$\La$ on $(C,\, S_{\mr{L}})$ plane corresponding to possible states
of the system and in particular identify maximally entangled mixed
states $\ro_{\mr{MEMS}}$ laying on the boundary of that subset. The
states $\ro_{\mr{MEMS}}$ have maximal allowed entanglement for a
given degree of mixedness. They obtained this result analytically
for some class of states. Numerical results suggests correctness of
the picture for general two-qubit states.
\par
In this note, we consider the problem of violation of Bell -- CHSH
inequalities \cite{Bell, CHSH} for mixed states. It is well known
that quantum states violating these inequalities have to be
entangled \cite{E, S}, but on the other hand, CHSH violation is
not necessary for mixed state entanglement \cite{Werner}. In the
context of the results of Ref.\cite{MJWK}, we address the
following question: what are the subsets of $\La$ which correspond
to states violating Bell -- CHSH inequalities ? In our previous
publication \cite{DJ}, we have studied the structure of such
subsets in the case of specific class of quantum states. The
results show that $\La$ is a sum of disjoint subsets
$\La_{\mr{V}},\, \La_{\mr{NV}}$ and $\La_{0}$ with the following
properties: states belonging to $\La_{\mr{V}}$ violate CHSH
inequalities, whereas states from $\La_{\mr{NV}}$  fulfil all CHSH
inequalities. The subset $\La_{0}$ has somehow unexpected
property: for any pair $(S_{\mr{L}},\, C)\in \La_{0}$ there are
two families of states with the same  entropy and concurrence such
that all states from one family violate CHSH and at the same time,
all states from the other family fulfil all CHSH inequalities. In
the present paper, we continue these investigations for general
class of two-qubit states. First we consider larger class of
states still admitting explicit formulas for linear entropy,
concurrence and degree of CHSH violation. Unfortunately,
analytical analysis of the relation between these functions is not
possible. Numerical investigations lead to some modifications of
the picture from Ref. \cite{DJ}, but the general structure is not
changed. Finally, this problem is studied using numerically
generated arbitrary density matrices. The results indicate that
the structure of $\La$ for general two-qubit states seems to be
the same.
\par
We consider also the problem of maximal violation of CHSH
inequalities. For a class of mixed states we obtain counterpart of
the result of Ref.\cite{MJWK}, namely we find the form of mixed
states with maximal degree of CHSH violation for given linear
entropy. All that states lie on specific curve on entropy --
concurrence plane. Numerical results suggest also that general two
-- qubits states with this property satisfy the same relation
between entropy and concurrence. As we show, maximal violation of
Bell inequalities with fixed linear entropy is not equivalent to the
maximal entanglement under the same conditions.
\section{CHSH inequalities}
Let $\tl{a},\,\tl{a}^{\prime},\,\tl{b},\,\tl{b}^{\prime}$ be the
unit vectors in $\R^{3}$ and
$\tl{\sigma}=(\si{1},\,\si{2},\,\si{3})$. Consider the family of
operators on two - qubits Hilbert space $\cH_{AB}=\C^{4}$
\begin{equation}
B_{CHSH}=\tl{a}\cdot\tl{\sigma}\otimes
(\tl{b}+\tl{b}^{\prime})\cdot\tl{\sigma}+\tl{a}^{\prime}\cdot\tl{\sigma}\otimes
(\tl{b}-\tl{b}^{\prime})\cdot\tl{\sigma}\label{B}
\end{equation}
Then Bell - CHSH  \cite{CHSH} inequalities are
\begin{equation}
|\tr (\ro\, B_{CHSH})|\leq 2\label{CHSH}
\end{equation}
If the above inequality is not satisfied by  the state $\ro$ for
some choice of $\tl{a},\,\tl{a}^{\prime},\, \tl{b},\,
\tl{b}^{\prime}$ , we say that $\ro$ \textit{violates Bell-CHSH
inequalities.} In the case of two-qubits, the violation of Bell -
CHSH inequalities by mixed states can be studied using simple
necessary and sufficient condition \cite{HHH,H}. Consider real
matrix
\begin{equation}
T_{\ro}=(t_{nm}),\quad t_{nm}=\tr
(\ro\,\si{n}\otimes\si{m})\label{Tmatix}
\end{equation}
and real symmetric matrix
\begin{equation}
U_{\ro}=T_{\ro}^{T}\,T_{\ro}\label{Umatix}
\end{equation}
where $T_{\ro}^{T}$ is the transposition of $T_{\ro}$. Let
\begin{equation}
m(\ro)=\max_{j<k}\; (u_{j}+u_{k})\label{m}
\end{equation}
and $u_{j},\, j=1,2,3$ are the eigenvalues of $U_{\ro}$.  As was
shown in \cite{HHH,H}
\begin{equation}
\max_{B_{CHSH}}\,\tr (\ro\, B_{CHSH})=2\,\sqrt{m(\ro)}
\end{equation}
Thus (\ref{CHSH}) is violated by some choice of
$\tl{a},\tl{a}^{\prime},\tl{b},\tl{b}^{\prime}$ if and only if
$m(\ro)>1$.
\par
We need also the measures of degree of entanglement and mixture for
given state. In the case of two qubits, the useful measure of degree
of entanglement is concurrence $C(\ro)$
\begin{equation}
 C(\ro)=\max\, (0,\,
2\lambda_{\mr{max}}(\widehat{\ro})-\tr
\widehat{\ro}\,)\label{concurrence}
\end{equation}
where $\lambda_{\mr{max}}(\widehat{\ro})$ is the maximal eigenvalue
of $\widehat{\ro}$ and
$$
\widehat{\ro}=\sqrt{\sqrt{\ro}\,\widetilde{\ro}\,\sqrt{\ro}},\quad
\widetilde{\ro}=(\si{2}\otimes\si{2})\overline{\ro}(\si{2}\otimes\si{2})
$$
with $\overline{\ro}$ denoting complex conjugation of the matrix
$\ro$. It is known that  $C(\ro)$ can be used to obtain entanglement
of formation, which is natural measure of entanglement for mixed
states \cite{HW,Woo}. To measure degree of mixture, or deviation
from pure state, we use linear entropy
\begin{equation}
S_{\mr{L}}(\ro)=\frac{4}{3}\,(1-\tr \ro^{2}\,)\label{linentr}
\end{equation}
which is normalized such that its maximal value equals $1$.
\section{CHSH inequalities and  entropy -- concurrence plane}
To study the subset of  entropy -- concurrence plane corresponding
to violation of CHSH inequalities, consider first the following
class $\fE_{0}$ of states
\begin{equation}
\ro=\begin{pmatrix} 0&0&0&0\\ 0&a&\frac{1}{2}ce^{i\theta}&0\\
0&\frac{1}{2}ce^{-i\theta}&b&0\\ 0&0&0&1-a-b
\end{pmatrix} \label{class}
\end{equation}
where
$$
c\in [0,1],\: a,b\geq 0,\: \theta\in [0,2\pi]
$$
and
$$
ab\geq\frac{c^{2}}{4},\quad a+b\leq 1
$$
Notice that for $\ro\in \fE_{0}$
$$
C(\ro)=c
$$
Define the subset $\La\subset \R^{2}$
\begin{equation}
\La=\{ (\,S_{L}(\ro),\,C(\ro)\,)\,:\, C(\ro)>0\quad\text{and}\quad
\ro\in \fE_{0}\}\label{obszar}
\end{equation}
In the paper \cite{DJ}, we  analysed the set (\ref{obszar}) and we
have shown that it is a sum of disjoint subsets $\La_{\mr{V}},\,
\La_{0}$ and $\La_{\mr{NV}}$ with properties:
\begin{enumerate}
\item[\textbf{1.}] If $(s,c)\in \La_{\mr{V}}$, then every state $\ro\in
\fE_{0}$ such that $S_{L}(\ro)=s$ and $C(\ro)=c$ satisfies
$m(\ro)>1$. \vskip 2mm\noindent
\item[ \textbf{2.}] If $(s,c)\in
\La_{0}$, then there exist states $\ro_{1},\, \ro_{2}\in \fE_{0}$
such that
$$
S_{L}(\ro_{1})=S_{L}(\ro_{2})=s,\quad C(\ro_{1})=C(\ro_{2})=c
$$
and $m(\ro_{1})>1$,  but $m(\ro_{2})<1$.
\vskip 2mm\noindent
\item[\textbf{3.}] If $(s,c)\in \La_{\mr{NV}}$, then every state $\ro\in
\fE_{0}$ such that $S_{L}(\ro)=s$ and $C(\ro)=c$ satisfies
$m(\ro)<1$.
\end{enumerate}
Detailed analytic description of regions $\La_{\mr{V}},\, \La_{0}$
and $\La_{\mr{NV}}$ as well as the proof of the above properties,
can be found in Ref. \cite{DJ}.
\par
In the present paper, we try to extend this analysis to the larger
class of two - qubit states. For general mixed two-qubit states
there is a bound on concurrence that guarantees violation of CHSH
inequalities irrespective of linear entropy. It follows from the
result of Verstraete and Wolf  \cite{VW} that minimal violation of
CHSH inequality for given concurrence $C$ is equal to
\begin{equation}
m_{\mr{min}}=\max \,(1,\, 2C^{2}\,)\label{minimal}
\end{equation}
So if
\begin{equation}
C(\ro)>\frac{1}{\sqrt{2}}\label{cbound}
\end{equation}
then minimal value of $m$ is greater then $1$, and every state
satisfying (\ref{cbound}) violates CHSH inequality. On the other
hand, there is a bound on linear entropy that guarantees fulfilling
CHSH inequalities irrespective of concurrence. It is given by the
result of Santos \cite{Santos} that all states with (normalized)
linear entropy
\begin{equation}
S_{\mr{L}}(\ro)>\frac{2}{3}\label{sbound}
\end{equation}
satisfy all CHSH inequalities. We see that these general bounds are
compatible with our previous analysis (FIG. 1) and possible
modifications can occur in region $\Lambda_{\mr{V}}$ below the line
$C=\frac{1}{\sqrt{2}}$ and in region $\Lambda_{\mr{NV}}$ on the left
hand side of the line $s=\frac{2}{3}$.
\begin{figure}[h]
\centering  {\includegraphics[height=56mm]{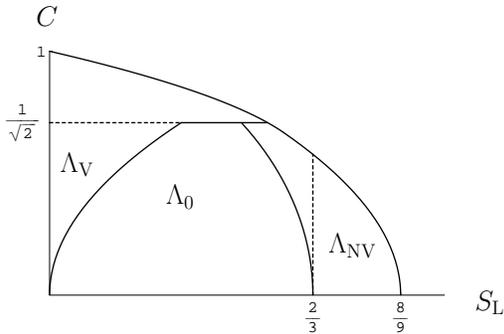}} \caption{The
structure of the set $\Lambda$ and Verstraete -- Wolf and Santos
bounds (dotted lines) }
\end{figure}
Consider now the larger class $\fE_{1}$ of states of the form
\begin{equation}
\ro=\begin{pmatrix} \ro_{11}&0&0&\ro_{14}\\
0&\ro_{22}&\ro_{23}&0\\
0&\ro_{32}&\ro_{33}&0\\
\ro_{41}&0&0&\ro_{44}
\end{pmatrix}\label{class1}
\end{equation}
One can check that for that class
\begin{equation}
C(\ro)=\max\, (0,\, C_{1},\, C_{2}\,)
\end{equation}
where
\begin{equation}
\begin{split}
&C_{1}=2\,(\, |\ro_{14}|-\sqrt{\ro_{22}\ro_{33}}\,)\\[2mm]
&C_{2}=2\, (\, |\ro_{23}|-\sqrt{\ro_{11}\ro_{44}}\,)
\end{split}
\end{equation}
and
\begin{equation}
S_{\mr{L}}(\ro)=1-\ro_{11}^2-\ro_{22}^2-\ro_{33}^2-\ro_{44}^2-2|\ro_{14}|^2-2|\ro_{23}|^2
\end{equation}
Moreover,
\begin{eqnarray}
m(\ro)=&&\max\,\big[\,4(|\ro_{41}|+|\ro_{23}|)^2,
4(|\ro_{41}|-|\ro_{23}|)^2, \nonumber\\
&&\hspace*{7mm}(\ro_{11}-\ro_{22}-\ro_{33}+\ro_{44})^2\,\big]
\end{eqnarray}
For the class (\ref{class1}), analytic description of regions of
$\Lambda$ where $m(\ro)>1$ or $m(\ro)<1$ is not possible, but it
can be done numerically. Results of numerical analysis of the
class $\fE_{1}$ are presented on FIG. 2. We see that the region
$\Lambda_{\mr{V}}$ where all states violate CHSH inequalities is
not changed, but there are states with $m(\ro)>1$ for some points
in $\Lambda_{\mr{NV}}$, so the region $\Lambda_{0}$ is slightly
enlarged.
\begin{figure}[h]
\centering  {\includegraphics[height=56mm]{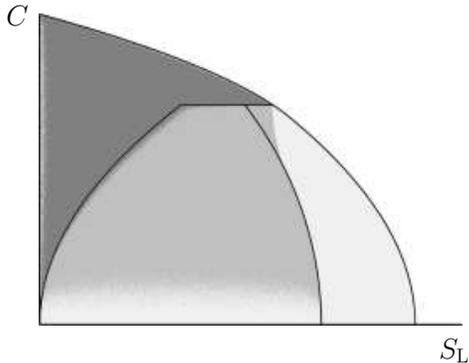}}
\caption{Numerical analysis of the set $\Lambda$ for the class
$\fE_{1}$: $m(\ro)>1$ (dark grey), $m(\ro)>1$ and $m(\ro)<1$ (grey),
$m(\ro)<1$ (light grey)}
\end{figure}
To study this problem for general two-qubit density matrices, we
numerically generate $3\cdot 10^{6}$ randomly chosen density
matrices. For such two-qubit states the structure of the set of
pairs $(S_{\mr{L}},C)$ is very simple. There are only points
corresponding to $m(\ro)>1$ or $m(\ro)<1$ (FIG. 3). Notice that by
the method of random choice of states, not all points of $\La$ are
achieved (the boundary of generated set correspond exactly to the
class of Werner states), but the obtained structure is compatible
with previous results. To have some insight into the properties of
the remaining part of the set $\La$, we modify the method of
generation of states and consider density matrices lying close to
the boundary of the set of all states i.e. such $\ro$ that one of
its eigenvalues is almost equal to zero. For these randomly
generated states, the pairs $(S_{\mr{L}},C)$ cover the whole set
$\La$, and its structure is the same as for the class $\fE_{1}$
(FIG. 4). The results suggest that the picture obtained using the
class $\fE_{1}$ should be correct also for all two-qubit density
matrices, although for most of randomly chosen density matrices
$\ro$ with fixed mixedness and linear entropy, either $m(\ro)>1$ or
$m(\ro)<1$. Unfortunately, we do not know analytic description of
the boundary of the enlarged region $\La_{0}$.
\begin{figure}[h]
\centering  {\includegraphics[height=56mm]{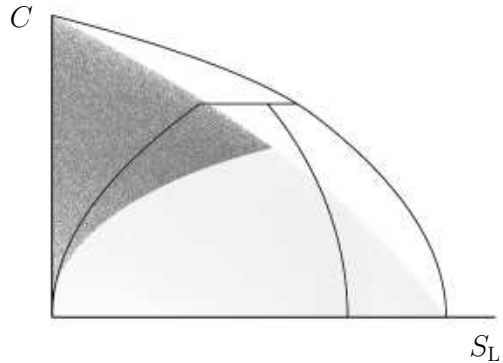}}
\caption{ The set $(S_{\mr{L}},\, C)$ for randomly chosen 
two-qubit states: $m(\ro)>1$ (dark grey), $m(\ro)<1$ (light grey).
The boundary corresponds to the family of Werner states}
\end{figure}
\begin{figure}[h]
\centering  {\includegraphics[height=56mm]{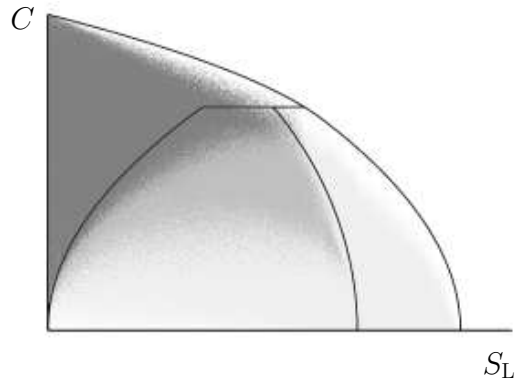}}
\caption{ The set $\Lambda$ for numerically generated states lying
close to the boundary: $m(\ro)>1$ (dark grey), $m(\ro)>1$ and
$m(\ro)<1$ (grey), $m(\ro)<1$ (light grey)}
\end{figure}
\section{Maximal CHSH violation}
Consider now the values of  $m(\ro)$ for states violating Bell --
CHSH inequalities. We are especially interested in maximal values
of $m(\ro)$. For the class (\ref{class})
\begin{equation}
m(\ro)=\max (\,2c^{2},\, (2(a+b)-1)^{2}+c^{2}\,)\label{m}
\end{equation}
We see that (\ref{m}) is maximal iff $a+b=1$, and then
\begin{equation}
m(\ro)=1+c^{2}\label{mmax}
\end{equation}
By general result of Verstraete and Wolf \cite{VW}, for any
two-qubit state, (\ref{mmax}) is the maximal degree of CHSH
violation for given concurrence $c$. But we ask  another question:
what is the maximum of  (\ref{mmax}) \textit{for fixed linear
entropy}, and which states realize that maximum? We can simply
answer this question for the class of states (\ref{class}). Since
\begin{equation}
S_{\mr{L}}(\ro)=\frac{4}{3}\,\left(1-a^{2}-(1-a)^{2}-\frac{c^{2}}{2}\,\right)\label{entropy}
\end{equation}
so fixing $S_{\mr{L}}(\ro)=s$, we obtain
\begin{equation}
m(\ro)=1+4\, (a-a^{2})-\frac{3}{2}s\label{msfixed}
\end{equation}
Maximum of (\ref{msfixed}) is achieved at $a=\frac{1}{2}$ and
equals to $2-\frac{\DS 3}{\DS 2}s$ for $s\in [0,\frac{2}{3}]$. In
this way we obtain
\begin{thm}
In the class (\ref{class}), states maximizing degree of violation of
CHSH inequalities for fixed linear entropy, lie on the curve
$$
s=\frac{2}{3}\,(1-c^{2}\,)
$$
and have the form
\begin{equation}
\ro_{\mr{MVB}}=\frac{1}{2}\begin{pmatrix}0&0&0&0\\
0&1&\sqrt{\beta-1}\,e^{i\theta}&0\\
0&\sqrt{\beta-1}\,e^{-i\theta}&1&0\\
0&0&0&0
\end{pmatrix}\label{mvb}
\end{equation}
with
$$
\beta\in [1,2],\: \theta\in [0,2\pi]
$$
Moreover,
$$
m(\ro_{\mr{MVB}})=\beta
$$
\end{thm}
\begin{figure}[h]
\centering  {\includegraphics[height=60mm]{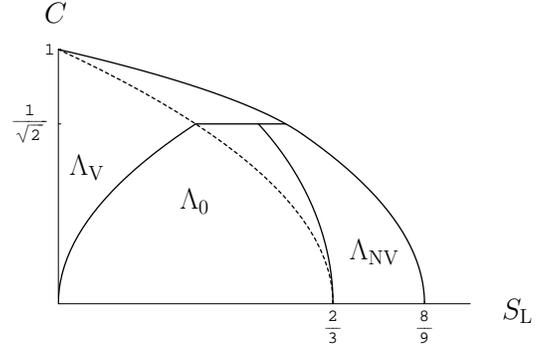}} \caption{
States with maximal degree of CHSH violation (dotted curve) on
$(S_{\mr{L}},\, C)$ plane}
\end{figure}
It is instructive to compare the value of $m(\ro_{\mr{MVB}})$ with
degree of CHSH violation for some other classes of states. Let $W$
be the family of Werner states
\begin{equation}
W=(1-p)\frac{\I_{4}}{4}+p\,\ket{\Psi^{-}}\bra{\Psi^{-}}\label{Werner}
\end{equation}
where $\Psi^{-}$ is a singlet state of two-qubits. Then
\begin{equation}
m(W)=2-2s\label{mW}
\end{equation}
For maximally entangled mixed states $\ro_{\mr{MEMS}}$ introduced in
Ref. \cite{MJWK}, the corresponding value of $m$ is given by
\begin{equation}
m(\ro_{\mr{MEMS}})=1-\frac{3}{4}s+\sqrt{1-\frac{3}{2}s}\label{mMEMS}
\end{equation}
We see that for a fixed linear entropy
\begin{equation}
m(\ro_{\mr{MVB}})\geq m(\ro_{\mr{MEMS}})\geq m(W)
\end{equation}
although
\begin{equation}
C(\ro_{\mr{MEMS}})\geq C(W)\geq C(\ro_{\mr{MVB}})
\end{equation}
So
\begin{cor}
The states with maximum amount of entanglement for a given linear
entropy do not maximize degree of Bell -- CHSH violation.
\end{cor}
\begin{figure}[ht]
\centering \rotatebox{270}
{\includegraphics[height=90mm]{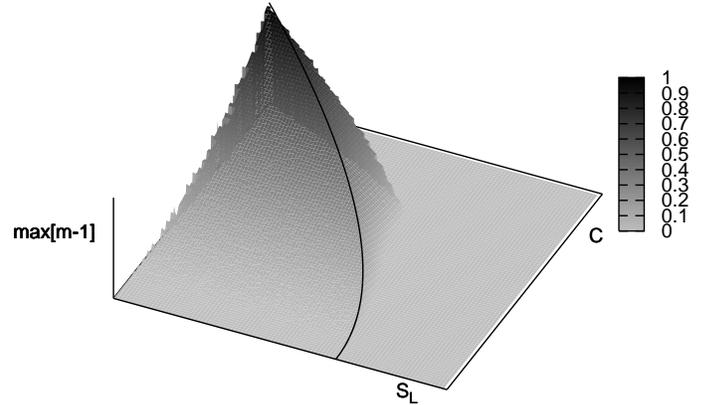}} \caption{ Plot of $\max\,
(m(\ro)-1\,)$ as the function of $S_{\mr{L}}$ and $C$ for the class
$\fE_{1}$. The curve corresponds to $m(\ro_{\mr{MVB}})-1$}
\end{figure}
The family of states with maximal degree of violation of Bell --
CHSH inequalities has another remarkable property. It is known that
fidelity of state $\ro$ defined as
\begin{equation}
F(\ro)=\max\, \ip{\psi}{\ro\psi}\label{fidelity}
\end{equation}
where the maximum is taken over all maximally entangled pure states
$\psi$ is bounded above by \cite{VV}
\begin{equation}
F(\ro)\leq \frac{1+C(\ro)}{2}\label{fidelity bound}
\end{equation}
By direct computation, one can check that
$$
F(\ro_{\mr{MVB}})=\frac{1+C(\ro_{\mr{MVB}})}{2}
$$
Thus
\begin{cor}
The states $\ro_{\mr{MVB}}$ maximize fidelity  for given
concurrence.
\end{cor}
For a larger class $\fE_{1}$ we have studied $m(\ro)$ as a
function of $S_{\mr{L}}$ and $C$ numerically. Again the results
agree with those obtained analytically for the class $\fE_{0}$
(FIG. 6).
\begin{acknowledgments}
L.J. acknowledges financial support by Polish Ministry of
Scientific Research and Information Technology under the grant
PBZ-Min-008/PO3/2003.
\end{acknowledgments}

\end{document}